\documentclass[10pt,twocolumn,letterpaper]{article}

\usepackage{cvpr}
\usepackage{times}
\usepackage{epsfig}
\usepackage{graphicx}
\usepackage{amsmath}
\usepackage{amssymb}
\usepackage[table]{xcolor}
\usepackage{subfigure}
\usepackage{epstopdf}
\usepackage[]{booktabs}
\usepackage{multirow}
\usepackage{siunitx}

\newcommand{\beforefigcaption}{\vspace{-4mm}}
\newcommand{\afterfigcaption}{\vspace{-4mm}}
\newcommand{\beforetab}{\vspace{-1mm}}
\newcommand{\aftertab}{\vspace{-4mm}}
\newcommand{\beforesection}{\vspace{-1mm}}
\newcommand{\aftersection}{\vspace{-1mm}}
\newcommand{\beforesubsection}{\vspace{0mm}}
\newcommand{\aftersubsection}{\vspace{0mm}}

\usepackage{etoolbox}
\newcommand{\zerodisplayskips}{%
  \setlength{\abovedisplayskip}{3pt}%
  \setlength{\belowdisplayskip}{3pt}%
  \setlength{\abovedisplayshortskip}{0pt}%
  \setlength{\belowdisplayshortskip}{0pt}}
\appto{\normalsize}{\zerodisplayskips}
\appto{\small}{\zerodisplayskips}
\appto{\footnotesize}{\zerodisplayskips}

\makeatletter
\newcommand{\thickhline}{%
    \noalign {\ifnum 0=`}\fi \hrule height 1pt
    \futurelet \reserved@a \@xhline
}
\makeatother

\usepackage[pagebackref=true,breaklinks=true,letterpaper=true,colorlinks,bookmarks=false]{hyperref}

\cvprfinalcopy 

\def\cvprPaperID{1506} 

\ifcvprfinal\fi

\begin{document}

\title{ARCH: Animatable Reconstruction of Clothed Humans}

\author{Zeng Huang$^{1,2}$\thanks{Work performed at Facebook Reality Labs.},\quad Yuanlu Xu$^{1}$,\quad Christoph Lassner$^{1}$,\quad Hao Li$^{2}$,\quad Tony Tung$^1$\\
$^1$Facebook Reality Labs, Sausalito, USA\quad\quad $^2$University of Southern California, USA\\
{\tt\small zenghuan@usc.edu, merayxu@gmail.com, classner@fb.com, hao@hao-li.com, tony.tung@fb.com}
}

\maketitle


\begin{abstract}
In this paper, we propose ARCH (Animatable Reconstruction of Clothed Humans), a novel end-to-end framework for accurate reconstruction of animation-ready 3D clothed humans from a monocular image.
Existing approaches to digitize 3D humans struggle to handle pose variations and recover details.
Also, they do not produce models that are animation ready.
In contrast, ARCH is a learned pose-aware model that produces detailed 3D rigged full-body human avatars from a single unconstrained RGB image.
A Semantic Space and a Semantic Deformation Field are created using a parametric 3D body estimator. They allow the transformation of 2D/3D clothed humans into a canonical space, reducing ambiguities in geometry caused by pose variations and occlusions in training data.
Detailed surface geometry and appearance are learned using an implicit function representation with spatial local features.
Furthermore, we propose additional per-pixel supervision on the 3D reconstruction using opacity-aware differentiable rendering.
Our experiments indicate that ARCH increases the fidelity of the reconstructed humans. We obtain more than 50\% lower reconstruction errors for standard metrics compared to state-of-the-art methods on public datasets. We also show numerous qualitative examples of animated, high-quality reconstructed avatars unseen in the literature so far.
\end{abstract}

\beforesection
\section{Introduction} \label{sec:intro}
\aftersection

3D human reconstruction has been explored for several decades in the field of computer vision and computer graphics.
Accurate methods based on stereo or fusion have been proposed using various types of sensors~\cite{furukawa07,tung09,tung12,newcombe15,xu2018scene,xu2016multi,xu2017multi}, and several applications have become popular in sports, medicine and entertainment (\eg, movies, games, AR/VR experiences).
However, these setups require tightly controlled environments.
To date, full 3D human reconstruction with detailed geometry and appearance
from in-the-wild pictures is still challenging (\ie, taken in natural conditions as opposed to laboratory environments).
Moreover, the lack of automatic rigging prevents animation-based applications.


Recent computer vision models have enabled the recovery of  2D and 3D human pose and shape estimation from a single image. 
However, they usually rely on representations that have limitations: (1) skeletons~\cite{Fang2018PoseGrammar} are kinematic structures that are accurate to represent 3D poses, but do not carry body shape information. (2) surface meshes~\cite{kanazawa2018hmr,NeuralBodyFit18,DenseRaCICCV19} can represent body shape geometry, but have topology constraints; (3) voxels~\cite{varol18_bodynet} are topology-free, but memory costly with limited resolution, and need to be rigged for animation.
In this paper, we propose the ARCH (Animatable Reconstruction of Clothed Humans) framework
that possesses all benefits of current representations. In particular, we introduce a learned model that has human body structure knowledge (i.e., body part semantics), and is trained with humans in arbitrary poses.

\begin{figure}[ptb]
\centering
\includegraphics[width=\linewidth]{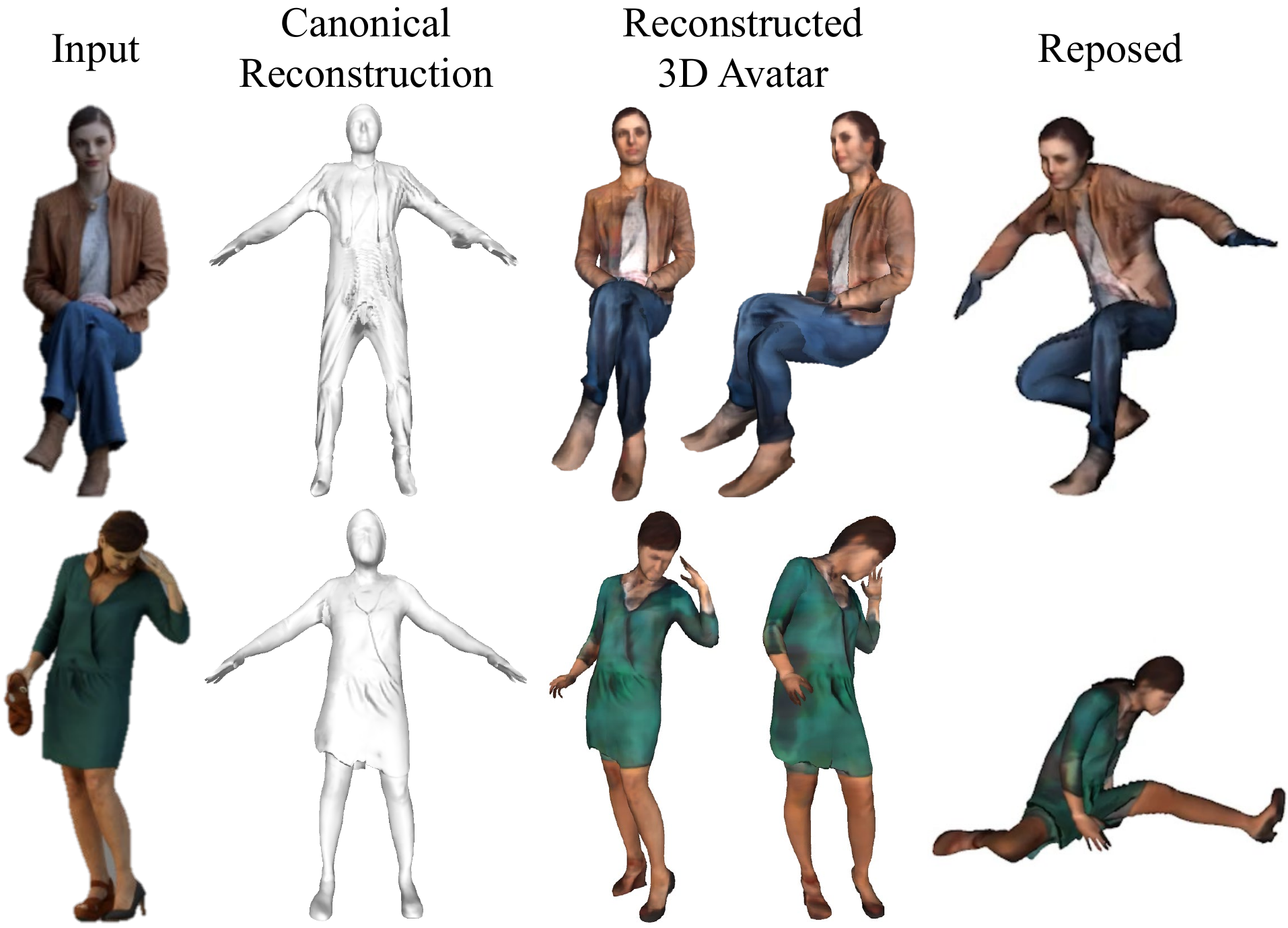}
\beforefigcaption
\caption{Given an image of a subject in arbitrary pose (left), ARCH creates an accurate and animatable avatar with detailed clothing (center). As rigging and albedo are estimated, the avatar can be reposed and relit in new environments (right).}
\afterfigcaption
\vspace{-2mm}
\label{fig:intro}
\end{figure}


\begin{figure*}[ptb]
\centering
\includegraphics[width=\textwidth]{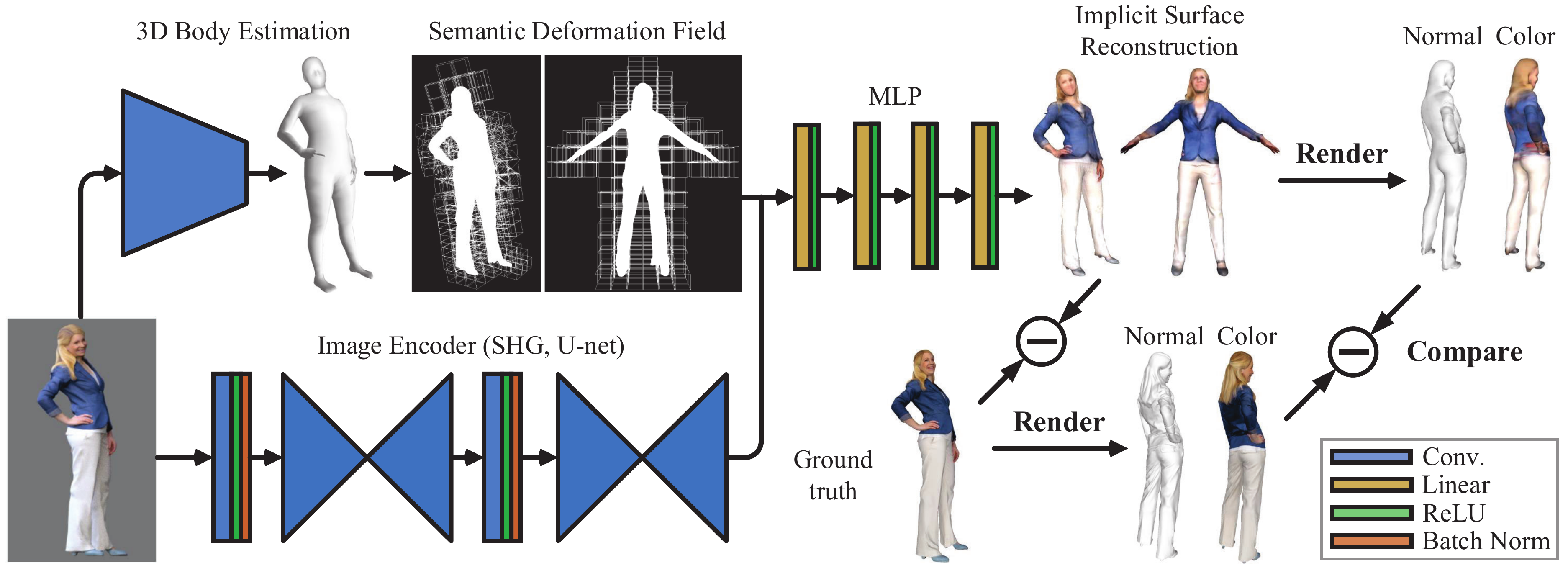}
\beforefigcaption
\caption{\emph{ARCH overview.} The framework contains three components: i) estimation of correspondences between an input image space and the canonical space, ii) implicit surface reconstruction in the canonical space from surface occupancy, normal and color estimation, iii) refinement of normal and color through differentiable rendering.}
\afterfigcaption
\label{fig:framework}
\end{figure*}

First, 3D body pose and shape estimation can be inferred from a single image of
a human in arbitrary pose by a prediction model~\cite{DenseRaCICCV19}. This initialization step is used for normalized-pose reconstruction of clothed human shape within a canonical space. This allows us to define a Semantic Space (SemS) and a Semantic Deformation Field (SemDF) by densely sampling 3D points around the clothed body surface and assigning skinning weights.
We then learn an implicit function representation of the 3D occupancy in the canonical space based on SemS and SemDF, which enables the reconstruction of high-frequency details of the surface (including clothing wrinkles, hair style, etc.) superior to the state of the art~\cite{SiCloPeCVPR19,PIFuICCV19,varol18_bodynet}. 
The surface representing a clothed human in a neutral pose is implicitly rigged in order to be used as an animatable avatar. 
Moreover, a differentiable renderer is used to refine normal and color information for each 3D point in space by Granular Render-and-Compare. Here, we regard them as a sphere and develop a new blending formulation based on the estimated occupancy.
See Fig.~\ref{fig:framework} for an overview of the framework.


In our experiments, we evaluate ARCH on the task of 3D human reconstruction from a single image.
Both quantitative and qualitative experimental results show ARCH outperforms state-of-the-art body reconstruction methods on public 3D scan benchmarks and in-the-wild 2D images. We also show that our reconstructed clothed humans can be animated by motion capture data, demonstrating the potential applications for human digitization for animation.

\textbf{Contributions}. The main contributions are threefold: 1)~we introduce the Semantic Space (SemS) and Semantic Deformation Field (SemDF) to handle implicit function representation of clothed humans in arbitrary poses, 2)~we propose opacity-aware differentiable rendering to refine our human representation via Granular Render-and-Compare, and 3)~we demonstrate how reconstructed avatars can directly be rigged and skinned for animation.
In addition, we learn per-pixel normals to obtain high-quality surface details, and surface albedo for relighting applications.

\beforesection
\section{Related Work} \label{sec:related}
\aftersection
\textbf{3D clothed human reconstruction} focuses on the task of reconstructing 3D humans with clothes. There are multiple attempts to solve this task with video inputs~\cite{Video3DAvatar3DV18,Video3DPeopleCVPR18,ClothCapTOG17,Video3DClothPeopleCVPR19,yangClothMocapECCV16}, RGB-D data~\cite{DoubleFusionCVPR18,HybridFusionECCV18} and in multi-view settings~\cite{MVDetailPoseShapeCVPR07,MVStereopsisTOG10,MVSkelSurfMocapCVPR09,MVArticulatedMeshAnimeTOG08,MVPhotoStereoTOG09,MVVisualHullTOG00,MVShadeMocapECCV12,bhatnagar2019mgn}. Though richer inputs clearly provide more information than single images, the developed pipelines yield more limitations on the hardware and additional time costs in deployment. Recently, some progress~\cite{bogo2016keep,HoloPoseCVPR19,kanazawa2018hmr,SPINICCV19,CMRCVPR19,lassner2017unite,tung2017self,DenseRaCICCV19,DeepHumanICCV19} has been made in estimating parametric human bodies from a single RGB image, yet boundaries are under-explored to what extent 3D clothing details can be reconstructed from such inputs.
In recent work~\cite{laehnerECCV18,lazova3dv2019,Tex2ShapeICCV19}, the authors learn to generate surface geometry details and appearance using 2D UV maps. While details can be learned, the methods cannot reconstruct loose clothing (\eg, dress) and recover complex shapes such as hair or fine structures (\eg, shoe heels).
Due to different types of clothing topology, volumetric reconstruction has great benefits in this scenario. For example, BodyNet~\cite{varol18_bodynet} takes a person image as input and learns to reconstruct voxels of the person with additional supervision through body priors (\eg, 2D pose, 3D pose, part mask); while PIFu~\cite{PIFuICCV19} assumes no body prior and learns an implicit surface function based on aligned image features, leading more clothes details and less robustness against pose variations.

In this paper, we incorporate body prior knowledge to transform people in arbitrary poses to the canonical space, and then learn to reconstruct an implicit representation.


\textbf{Differentiable rendering} makes the rendering operation differentiable and uses it to optimize parameters of the scene representation. Existing approaches can be roughly divided into two categories: mesh rasterization based rendering~\cite{chen2019dibrender,kato2018renderer,liu2019softras,loper2014opendr,LiDRRayTrace18} and volume based rendering~\cite{EldarDRPointCloud18,liu2019ImplicitSurfDR}. For example, OpenDR~\cite{loper2014opendr} and Neural Mesh Renderer~\cite{kato2018renderer} manually define approximated gradients of the rendering operation to move the faces. SoftRasterizer~\cite{liu2019softras} and DIB-R~\cite{chen2019dibrender}, in contrast, redefine the rasterization as a continuous and differentiable function, allowing gradients to be computed automatically. For volume-based differentiable rendering, \cite{EldarDRPointCloud18} represents each 3D point as a multivariate Gaussian and performs occlusion reasoning with grid discretization and ray tracing. Such methods require an explicit volume to perform occlusion reasoning. \cite{liu2019ImplicitSurfDR} develops differentiable rendering for implicit surface representations with a focus on reconstructing rigid objects.

In contrast, we use a continuous rendering function as in \cite{liu2019softras}, but revisit it to handle opacity, and we use geometric primitives at points of interest and optimize their properties.

\beforesection
\section{Proposed Framework} \label{sec:method}
\aftersection

ARCH contains three components, after 3D body estimation by~\cite{DenseRaCICCV19} (see Fig.~\ref{fig:framework}):
pose-normalization using Semantic Space (SemS) and Semantic Deformation Field (SemDF), implicit surface reconstruction, and refinement using a differentiable renderer by Granular Render-and-Compare (see Sec.~\ref{sec:rac}).

\beforesubsection
\subsection{Semantic Space and Deformation Field} \label{sec:bdf}
\aftersubsection

Our goal is to transform an arbitrary (deformable) object into a \textit{canonical space} where the object is in a predefined \textit{rest pose}.
To do so, we introduce two concepts: the Semantic Space (SemS) and the Semantic Deformation Field (SemDF).
SemS $S = \{(p, s_p): p\in\mathbb{R}^3\}$ is a space consisting of 3D points where each point $p \in S$ is associated to semantic information $s_p$ enabling the transformation operation. SemDF is a vector field represented by a vector-valued function $\mathcal{V}$ that accomplishes the transformation,

In computer vision and graphics, 3D human models have been widely represented by a kinematic structure mimicking the anatomy that serves to control the pose, and a surface mesh that represents the human shape and geometry. Skinning is the transformation that deforms the surface
given the pose. It is parameterized by skinning weights that individually influence body part transformations~\cite{loper2015smpl}.
In ARCH, we define SemS in a similar form, with skinning weights.

Assuming a skinned body template model $T$ in a normalized A-pose (i.e., the \textit{rest pose}), its associated skeleton in the canonical space, and skinning weights $W$, SemS is then
\begin{equation} \small
    S = \{(p, \{w_{i,p}\}_{i=1}^{N_K}): p \in \mathbb{R}^3\},
\end{equation}
where each point $p$ is associated to a collection of skinning weights $\{w_{i,p}\}$ defined with respect to $N_K$ body parts (e.g., skeleton bones). In this paper, we approximate $\{w_{i,p}\}$ by retrieving the closest point $p'$ on the template surface to $p$ and assigning the corresponding skinning weights from $W$. In practice, we set a distance threshold to cut off points that are too far away from $T$.

In ARCH, SemDF actually performs an \textit{inverse-skinning} transformation, putting a human in arbitrary pose to its normalized-pose in the canonical space.
This extends standard skinning (e.g., Linear Blend Skinning or LBS~\cite{loper2015smpl}) applied to structured objects to arbitrary 3D space and enables transforming an entire space in arbitrary poses to the canonical space, as every point $p'$ can be expressed as a linear combination of points $p$ with skinning weights $\{w_{i,p}\}$.

Following LBS, the canonical space of human body is tied to a skeletal rig. The state of the rig is described by relative rotations $R = \{r_i\}_{i=1}^{N_K}$ of all skeleton joints $X = \{x_i\}_{i=1}^{N_K}$. Every rotation is relative to the orientation of the parent element in a kinematic tree. For a skeleton with $N_K$ body parts, $R \in \mathbb{R}^{3 \times N_K}, X \in \mathbb{R}^{3 \times N_K}$.
Given a body template model $T$ in rest pose with $N_V$ vertices, the LBS function $\mathcal{V}(v_i, X, R; W)$ takes as input the vertices $v_i \in T$, the joints $X$, a target pose $R$,
and deforms every $v_i$ to the posed position $v_i'$ with skinning weights $W \in \mathbb{R}^{N_V \times N_K}$, namely,
\begin{equation}\small \label{eq:lbs}
    \mathcal{V}(v_i, X, R; W) = \sum\nolimits_{k=1}^{N_K} w_{k,i}\, G_k(R, X)\, v_i,
\end{equation}
where $G_k(R, X)$ is the rest-pose corrected affine transformation to apply to body part $k$.

\beforesubsection
\subsection{Implicit Surface Reconstruction} \label{sec:model}
\aftersubsection

We use the occupancy map $O$ to implicitly represent the 3D clothed human, \ie, 
\begin{equation}\small
    O = \{ (p, o_p):\; p \in \mathbb{R}^3,\, 0 \leq o_p \leq 1\}, 
\end{equation}
where $o_p$ denotes the occupancy for a point $p$. To obtain a surface, we can simply threshold $\tau$ the occupancy map $O$ to obtain the isosurface $O'_\tau$.

In this paper, we incorporate a human body prior by always reconstructing a neutral-posed shape in the canonical space.
Similar to~\cite{PIFuICCV19}, we develop a deep neural network that takes a canonical space point $p$, its correspondent 2D position $q$, and the 2D image $I$ as inputs and estimates occupancy $o_p$, normal $n_p$, color $c_p$ for $p$; that is,
\begin{equation} \small \begin{aligned}
o_p &= \mathcal{F}(f_p^s, I; \theta_o), \\
n_p &= \mathcal{F}(f_p^s, I, f_p^o; \theta_n), \\
c_p &= \mathcal{F}(f_p^s, I, f_p^o, f_p^n; \theta_c), \\
f_p^s \in \mathbb{R}^{171},\; & f_p^o \in \mathbb{R}^{256},\; f_p^n \in \mathbb{R}^{64},\; f_p^c \in \mathbb{R}^{64},
\end{aligned} \end{equation}
where $\theta^o$, $\theta^n$ and $\theta^c$ denote the occupancy, normal and color sub-network weights, $f_p^s$ is the \emph{spatial feature} extracted based on SemS. We use the estimated $57$ canonical body landmarks from~\cite{DenseRaCICCV19} and compute the Radial Basis Function (RBF) distance between $p$ and the $i$-th landmark $p_i'$, that is
\begin{equation} \small
f_p^s(i) = \exp\{-\mathcal{D}(p, p_i')\},
\end{equation}
where $\mathcal{D}(\cdot)$ is the Euclidean distance. We also evaluate the effects of different types of spatial features in Sec.~\ref{sec:result}.
$f_p^o$ and $f_p^n$ the feature maps extracted from occupancy and normal sub-networks, respectively (see also Fig.~\ref{fig:framework}). The three sub-networks are defined as follows:

The \textbf{Occupancy sub-network} uses a Stacked Hourglass (SHG)~\cite{newell2016stacked} as the image feature encoder and a Multi-Layer Perceptron (MLP) as the regressor. Given a $512 \times 512$ input image $I$, the SHG produces a feature map $f \in \mathbb{R}^{512 \times 512 \times 256}$ with the same grid size. For each 3D point $p$, we consider the feature located at the corresponding projected pixel $q$ as its visual feature descriptor $f_p^o \in \mathbb{R}^{256}$. For points that do not align onto the grid, we apply bi-linear interpolation on the feature map to obtain the feature at that pixel-aligned location.
The MLP takes the spatial feature of the 3D point $p \in \mathbb{R}^3$ and the pixel-aligned image features $f_p^o \in \mathbb{R}^{256}$ as inputs and estimates the occupancy $o_p \in [0, 1]$ by classifying whether this point lies inside the clothed body or not. 

The \textbf{Normal sub-network} uses a U-net~\cite{U-Net} as the image feature encoder and a MLP which takes the spatial feature, and feature descriptors $f_p^n \in \mathbb{R}^{64}$ and $f_p^o \in \mathbb{R}^{256}$ from its own backbone and from the occupancy sub-network as inputs and estimates the normal vector $n_p$.

The \textbf{Color sub-network} also uses a U-net~\cite{U-Net} as the image feature encoder and a MLP which takes the spatial feature, and feature descriptors $f_p^c \in \mathbb{R}^{64}$, $f_p^n \in \mathbb{R}^{64}$ and $f_p^o \in \mathbb{R}^{256}$ from its own backbone, as well as the normal and occupancy sub-networks as inputs and estimates the color $c_p$ in RGB space.

For each sub-network, the MLP takes the pixel-aligned image features and the spatial features (as described in Sec.~\ref{sec:bdf}), where the numbers of hidden neurons are $(1024,512,256,128)$. Similar to~\cite{PIFuICCV19}, each layer of MLP has skip connections from the input features.
For the occupancy sub-network, the MLP estimates one-dimension occupancy $o_p \in [0, 1]$ using Sigmoid activation.
For the normal sub-network, the MLP estimates three-dimension normal $n_p \in [0, 1]^3, \left\lVert n_p\right\rVert_2 = 1$ using L2 normalization.
For the color sub-network, the MLP estimates three-dimension color $c_p \in [0, 1]^3$ using range clamping.


\begin{figure*}
\centering
\includegraphics[width=\textwidth]{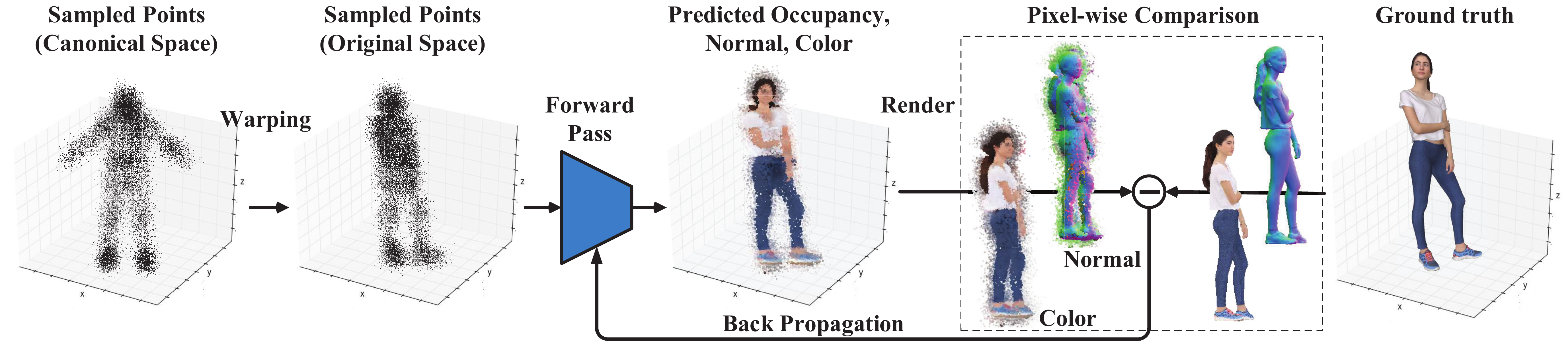}
\beforefigcaption
\caption{\textit{Illustration of the loss computation through differentiable rendering}. From left to right: points are sampled according to a Gaussian distribution around our template mesh in the canonical space. They are transformed with the estimated Semantic Deformation Field and processed by the model. The model provides estimations of occupancy, normal and color for each 3D point. We use a differentiable renderer to project those points onto a new camera view and calculate pixel-wise differences to the rendered ground truth.}
\afterfigcaption
\label{fig:rac}
\end{figure*}

\beforesubsection
\subsection{Training} \label{sec:learning}
\aftersubsection

During training, we optimize the parameters of all three sub-models, \ie, the occupancy, normal and color models. We define the training in three separate loops to train each part with the appropriate losses and avoid computational bottlenecks. The total loss function is defined as
\begin{equation}\small
\mathcal{L} = \mathcal{L}_{3d}^o + \mathcal{L}_{3d}^n + \mathcal{L}_{3d}^c + \mathcal{L}_{2d}^n + \mathcal{L}_{2d}^c,
\end{equation}
where $\mathcal{L}_{3d}^o$ is the 3D loss for occupancy network, $\mathcal{L}_{3d}^n$ and $\mathcal{L}_{2d}^n$ are the 3D and 2D losses for normal network, and $\mathcal{L}_{3d}^c$ and $\mathcal{L}_{2d}^c$ are the 3D and 2D losses for color network. For every training iteration, we perform the following three optimizations.

\textbf{Occupancy}. We use the available ground truth to train the occupancy prediction model in a direct and supervised way. First, we sample \num{20480} points in the canonical space. They are sampled around the template mesh according to a normal distribution with a standard deviation of \SI{5}{cm}. This turned out to cover the various body shapes and clothing well in our experiments, but can be selected according to the data distribution at hand. These points are then processed by the occupancy model, providing us with an estimated occupancy value for every sampled point. We use a \textit{sigmoid} function on these values to normalize the network output to the interval $[0, 1]$, where we select \num{0.5} as the position of the isosurface. \num{0.5} is the position where the derivative of the \textit{sigmoid} function is the highest and we expect to optimize the surface prediction best. The loss $\mathcal{L}_{3d}^o$ is defined as the Huber loss comparing the occupancy prediction and ground truth. Similar to~\cite{Park_2019_CVPR}, we found a less aggressive loss function than the squared error better suited for the optimization, but found the quadratic behavior of the Huber loss around zero to be beneficial.

\textbf{Normals and colors for surface points}. Colors and normals can be optimized directly from the ground truth mesh for points that lie on its surface. To use this strong supervision signal we introduce a dedicated training stage. In this stage, we sample points only from the mesh surface and push them through the color and normal models. In our setup, we use \num{51200} point samples per model per training step. The loss terms $\mathcal{L}_{3d}^n$ and $\mathcal{L}_{3d}^c$ are defined as the $L1$ loss comparing the predicted normals and colors with the ground truth across all surface points. The occupancy predictions are kept unchanged.

\textbf{Normals and colors for points not on the surface}. For points not on the mesh surface, it is not clear how the ground truth information can be used in the best way to improve the prediction without an additional mapping. In a third step for the training, we sample another set of \num{51200} points, and push them through the occupancy, color and normal models and use a differentiable renderer on the prediction. We render the image using the occupancy information as opacity, and by using the color channels to represent colors or normals and use the gradients to update the predicted values. $\mathcal{L}_{2d}^n$ and $\mathcal{L}_{2d}^c$ are defined as the per-pixel L1 loss between the rendered image and the ground truth. For details on this step, see Fig.~\ref{fig:rac} and the following Sec.~\ref{sec:rac}.

\beforesubsection
\subsection{Granular Render-and-Compare} \label{sec:rac}
\aftersubsection

The prediction from the model is an implicit function representation.
By sampling points in a predefined volume and optimizing $\mathcal{L}_{3d}^o$, $\mathcal{L}_{3d}^n$ and $\mathcal{L}_{3d}^c$, we can optimize the occupancy, normal and color at these points directly given 3D ground truth. However, it is not clear what the gradients should be for points that are located not directly on the surface of the ground truth mesh. To address this problem, we propose to use a differentiable renderer.

We first create an explicit geometric representation of the scene at hand.
For every sample point to optimize, we place a geometric primitive with a spatial extent at its position. 
To be independent of the viewpoint, we choose this to a sphere with \SI{1}{cm} radius for every sampled point (for an overview of the differentiable rendering loss computation, see Fig.~\ref{fig:rac}). During training, every scene to render contains \num{51200} spheres.

We then define a differentiable rendering function~\cite{liu2019softras} to project the spheres onto the image plane so that we can perform pixel-level comparisons with the projected ground truth. We use a linear combination with a weight $w^i_j$ to associate the color contribution from point $p_i$ to the pixel $q_j$. Having the color $c_i$ and normal $n_i$ for point $p_i$, the color and normal for pixel $q_j$ are calculated as the weighed linear combination of point values $\sum\nolimits_i w^i_j c_i$ and $\sum\nolimits_i w^i_j n_i$.

We define $w^i_j$ considering two factors: the depth of the sphere for point $p_i$ at pixel $q_j$, $z^i_j$, and the proximity of the projected surface of the sphere for point $p_i$ to pixel $q_j$, $d^i_j$. To make occlusion possible, the depth needs to have a strong effect on the resulting weight. Hence, \cite{liu2019softras} defines the weight as
\begin{equation} \small \label{eq:weight_plain}
w^i_j=\frac{d^i_j \exp(z^i_j/\gamma)}{\sum_k d^i_k\exp(z^i_k/\gamma)+\exp(\epsilon/\gamma)}
\end{equation}
with $\epsilon$ being a small numerical constant. With this definition, the proximity has linear influence on the resulting weight while the depth has exponential influence. The impact ratio is controlled by the scaling factor $\gamma$, which we fix to \num{1e-5} in our experiments.

In contrast to \cite{liu2019softras} we also need to use an opacity $\alpha_i$ per sphere for rendering. We tie this opacity value $\alpha_i$ directly to the predicted occupancy value through linear scaling and shifting. To stay with the formulation of the render function, we integrate $\alpha_i$ into the weight formulation in Eqn.~\ref{eq:weight_plain}.

If the opacity is used as a linear factor in this equation, the \textit{softmax} function will still render spheres with very low opacity over other spheres with a lower depth value. The problem is the exponential function that is applied to the scaled depth values. On the other hand, if an opacity factor is only incorporated into the exponential function, spheres will remain visible in front of the background (their weight factor is still larger than the background factor $\exp(\epsilon/\gamma)$). We found a solution by using the opacity value as both, linear scaling factor as well as exponential depth scaling factor. This solution turned out to be numerically stable and well-usable for optimization with all desired properties. This changes the weight function to the following:
\begin{equation} \small
w^i_j=\frac{\alpha^i d^i_j \exp(\alpha^i z^i_j/\gamma)}{\sum_k \alpha^i d^i_k\exp(\alpha^i z^i_k/\gamma)+\exp(\epsilon/\gamma)}.
\end{equation}

Using this formulation, we optimize the color channel values $c_i$ and normal values $n_i$ per point. A per-pixel L1 loss is computed between the rendering and a rendering of the ground truth data and back-propagated through the model. For our experiments with $\gamma=\num{1e-5}$ and the depth of the volume, we map the occupancy values that define the isosurface at the value \num{0.5} to the threshold where $\alpha$ shifts to transparency. We experimentally determined this value to be roughly~\num{0.7}.

\beforesubsection
\subsection{Inference}
\aftersubsection

For inference, we take as input a single RGB image representing a human in an arbitrary pose, and run the forward model as described
in Sec.~\ref{sec:model} and Fig.~\ref{fig:framework}.
The network outputs a densely sampled occupancy field over the canonical space from which we use the Marching Cube algorithm~\cite{MarchCube87} to extract the isosurface at threshold $0.5$.
The isosurface represents the reconstructed clothed human in the canonical pose.
Colors and normals for the whole surface are also inferred by the forward pass and are pixel-aligned to the input image (see Sec.~\ref{sec:model}).
The human model can then be transformed to its original pose $R$ by LBS using SemDF and per-point corresponding skinning weights $W$ as defined in Sec.~\ref{sec:bdf}.

Furthermore, since the implicit function representation is equipped with skinning weights and skeleton rig, it can naturally be warped to arbitrary poses.
The proposed end-to-end framework can then be used to create a detailed 3D avatar that can be animated with unseen sequences from a single unconstrained photo (see Fig.~\ref{fig:animation}).

\beforesection
\section{Experiments} \label{sec:experiments}
\aftersection

We present details on ARCH implementation and datasets for training, with results and comparisons to the state of the art.

\beforesubsection
\subsection{Implementation Details}
\aftersection


ARCH is implemented in PyTorch. We train the neural network model using the RMSprop optimizer with a learning rate starting from 1e-3. The learning rate is updated using an exponential schedule every 3 epochs by multiplying with the factor \num{0.1}. We are using \num{582} 3D scans to train the model and use \num{360} views per epoch, resulting in \num{209520} images for the training per epoch. Training the model on an NVIDIA DGX-1 system with one Tesla V100 GPU takes \SI{90}{h} for 9 epochs.

\beforesubsection
\subsection{Datasets}
\aftersection

\begin{figure}
\centering
\includegraphics[width=0.95\linewidth]{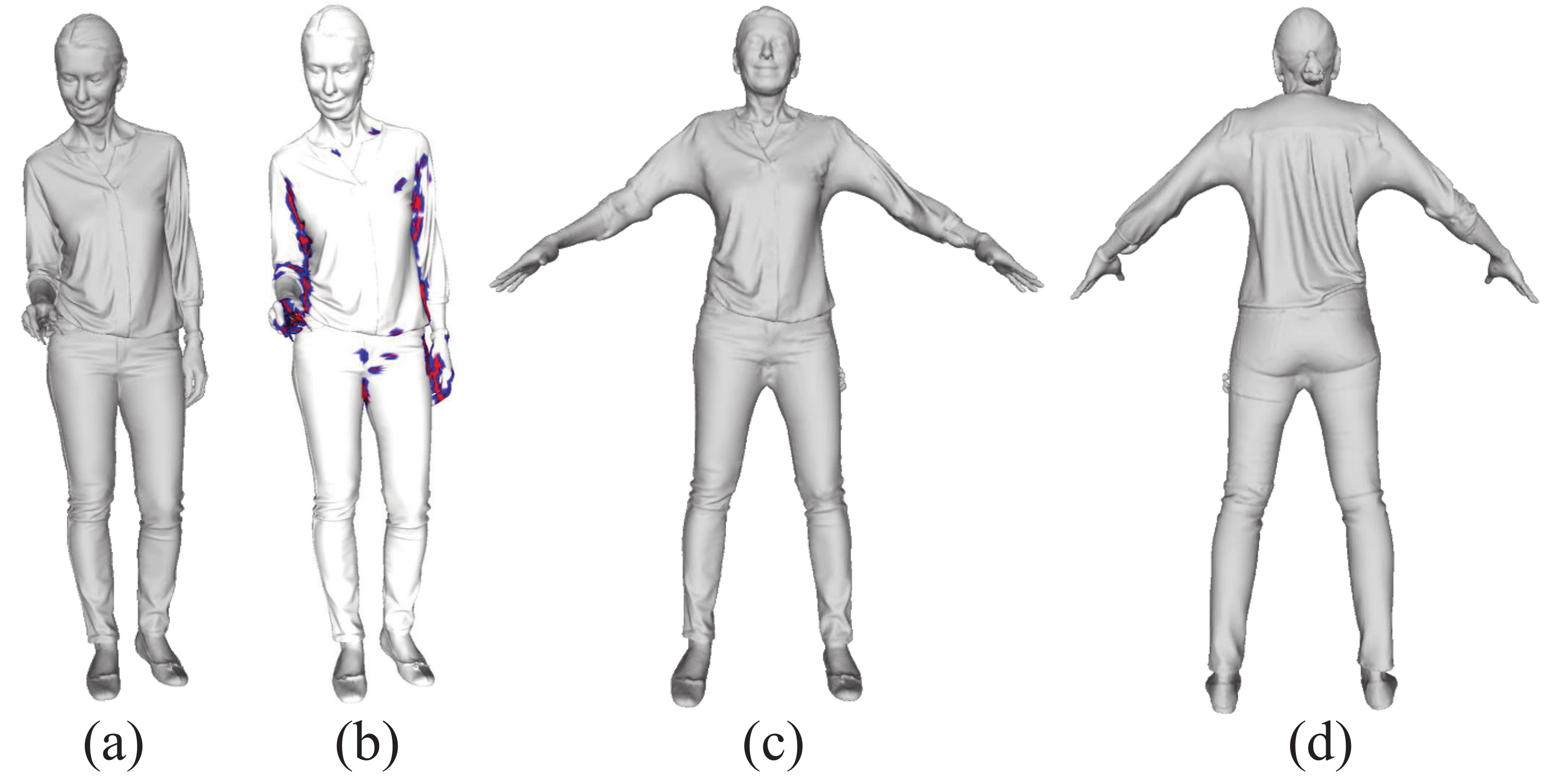}
\beforefigcaption
\vspace{3mm}
\caption{\textit{Illustration of reposing 3D scans to the canonical space.} \textbf{(a)} An original 3D scan from the RenderPeople dataset. \textbf{(b)} Automatically detected topology changes. Red marks points with self contacts, blue regions that are also removed before reposing to avoid problems with normals. \textbf{(c, d)} Reposed scan.}
\afterfigcaption
\label{fig:repose}
\end{figure}

Our training dataset is composed of 375 3D scans from the RenderPeople\footnote{\url{http://renderpeople.com}} dataset, and 207 3D scans from the AXYZ\footnote{\url{http://secure.axyz-design.com}} dataset. The scans are watertight meshes which are mostly free of noise.
They represent subjects wearing casual clothes, and potentially holding small objects (\eg, mobile phones, books and purses).
Our test dataset contains 64 scans from the RenderPeople dataset, 207 scans from the AXYZ dataset, 26 scans from the BUFF dataset~\cite{zhang-CVPR17}, and 2D images from the DeepFashion~\cite{liu2016deepfashion} dataset, representing clothed people with a large variety of complex clothing.
The subjects in the training dataset are mostly in standing pose, while the subjects in the test dataset are in arbitrary poses (standing, bending, sitting, \ldots).
We create renders of the 3D scans using Blender.
For each 3D scan, we produce 360 images by rotating a camera around the vertical axis with intervals of 1 degree.
For the current experiments, we only considered the weak perspective projection (orthographic camera) but this can be easily adapted.
We also used 38 environment maps to render each scan with different natural lighting conditions.
The proposed model is trained to predict albedo (given by ground truth scan color).
We also observed that increasing the number of images improves the fidelity of predicted colors (as in~\cite{PIFuICCV19}).

In order to use a 3D scan for model training, we fit a rigged 3D body template to the scan mesh to estimate the 3D body pose (see Fig.~\ref{fig:repose}).
The estimated parametric 3D body can directly serve as ground truth input data during the model training step (see Sec.~\ref{sec:learning}).
This also allows us to obtain SemS and SemDF for the scan.
However, since each 3D scan has its own topology, artifacts due to topology changes will occur when pose-normalization is naively applied to models containing self-contact (for example arms touching the body). This creates inaccurate deformations.
Hence, we first detect regions of self-contact and topology changes and cut the mesh before pose-normalization (see Fig.~\ref{fig:repose} (c) and (d)). Holes are then filled up using Smooth Signed Distance Surface reconstruction~\cite{calakli11}  (see Fig.~\ref{fig:repose} (c) and (d)).
For inference on 2D images from the DeepFashion dataset, we obtain 3D body poses using the pre-trained models from~\cite{DenseRaCICCV19}.

\beforesubsection
\subsection{Results and Comparisons} \label{sec:result}
\aftersection

\begin{table}[ptb]
\centering
\setlength{\tabcolsep}{6pt}
\renewcommand\arraystretch{1.1}
\resizebox{\linewidth}{!}{
\begin{tabular}{@{}r|c|c|c||c|c|c@{}}
\hline\thickhline
\multirow{2}{*}{Methods}  & \multicolumn{3}{c||}{\textbf{RenderPeople}} & \multicolumn{3}{c}{\textbf{BUFF}} \\
\cline{2-7}                             & Normal    & P2S     & Chamfer    & Normal  & P2S         & Chamfer \\ 
\thickhline
BodyNet~\cite{varol18_bodynet}          & 0.26      & 5.72    & 5.64     & 0.31     & 4.94    & 4.52   \\ 
SiCloPe~\cite{SiCloPeCVPR19}    	    & 0.22      & 3.81    & 4.02     & 0.22     & 4.06    & 3.99   \\ 
IM-GAN~\cite{chen2018implicit_decoder}  & 0.26      & 2.87    & 3.14     & 0.34     & 5.11    & 5.32   \\ 
VRN~\cite{VolumeRegECCVW2018}        	& 0.12      & 1.42    & 1.6      & 0.13     & 2.33    & 2.48   \\ 
PIFu~\cite{PIFuICCV19}      		    & 0.08      & 1.52    & 1.50     & 0.09     & 1.15    & 1.14   \\ 
\thickhline
ARCH, baseline   & 0.080 & 1.98 & 1.85 & 0.081 & 1.74 & 1.75 \\
+ SemDF      		& 0.042 & \textbf{0.74} & \textbf{0.85} & 0.045 & \textbf{0.82} & \textbf{0.87} \\
+ GRaC          & \textbf{0.038} & \textbf{0.74} & \textbf{0.85} & \textbf{0.040} & \textbf{0.82} & \textbf{0.87} \\
\hline\thickhline
\end{tabular}}
\beforetab
\caption{\textit{Quantitative comparisons of normal, P2S and Chamfer errors between posed reconstruction and ground truth on the RenderPeople and BUFF datasets.} Lower values are better.}
\aftertab
\label{tab:rec}
\end{table}

\begin{figure*}
\centering
\includegraphics[width=\textwidth]{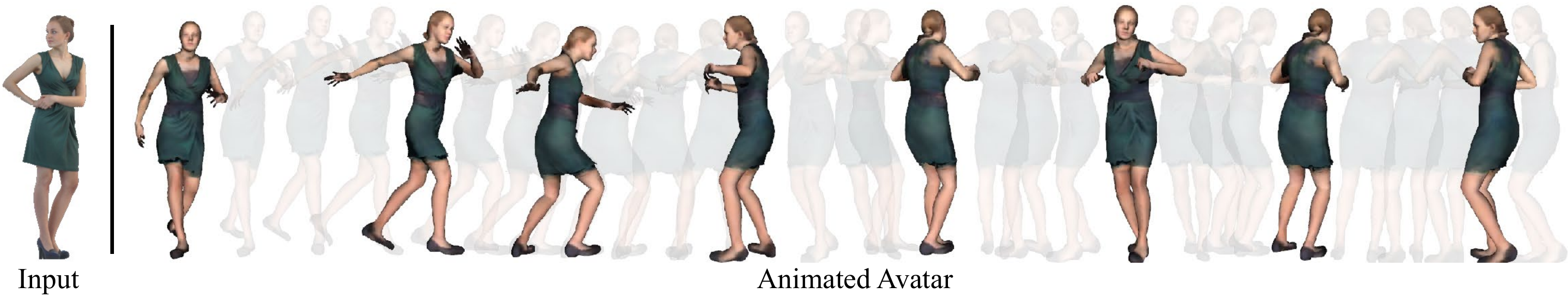}
\beforefigcaption
\caption{\textit{An example for animating a predicted avatar.} We use a predicted, skinned avatar from our test set and drive it using off-the-shelf motion capture data. This avatar has been created using only a single, frontal view. Our model produces a plausible prediction for the unseen parts, for example the hair and the back of the dress.}
\afterfigcaption
\label{fig:animation}
\end{figure*}

We evaluate the reconstruction accuracy of ARCH with three metrics similar to~\cite{PIFuICCV19}.
We reconstruct the results on the same test set and repose them back to the original poses of the input images and
compare the reconstructions with the ground truth surfaces in the original poses.
We report the average point-to-surface Euclidean distance (P2S) in centimeters, the Chamfer distance in centimeters,
and the L2 normal re-projection error in Tab.~\ref{tab:rec}.

Additionally to comparing with state-of-the-art methods~\cite{chen2018implicit_decoder,VolumeRegECCVW2018,kanazawa2018hmr,SiCloPeCVPR19,PIFuICCV19,varol18_bodynet}, we include scores of an ablative study with the proposed method. In particular, we evaluate three variants and validate the effectiveness of two main components: the Semantic Deformation Field and the Granular Render-and-Compare loss.

\textit{ARCH, baseline}: a variant of~\cite{PIFuICCV19} using our own network specifications, taking an image as input and directly estimating the implicit surface reconstruction. 

\textit{Semantic Deformation Field (SemDF)}: we first estimate the human body configuration by~\cite{DenseRaCICCV19} and then reconstruct the canonical shape using the implicit surface reconstruction, and finally repose the canonical shape to the original pose in the input image.

\textit{Granular Render-and-Compare (GRaC)}: based on the previous step, we further refine the reconstructed surface normal and color using differentiable render-and-compare.

\begin{figure}
\centering
\includegraphics[width=\linewidth]{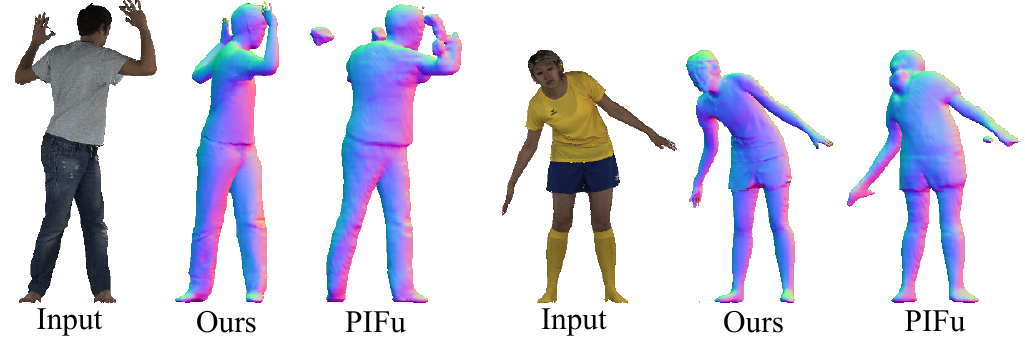}
\beforefigcaption
\caption{\textit{Evaluation on BUFF.} Our method outperforms~\cite{PIFuICCV19} for detailed reconstruction from arbitrary poses. We show results from different angles.}
\vspace{2mm}
\afterfigcaption
\label{fig:BUFF}
\end{figure}

ARCH baseline specification already achieves state-of-the-art performance in normal estimation, but has inferior performance w.r.t. P2S and Chamfer error compared to PIFu~\cite{PIFuICCV19}. We use a different training dataset compared to PIFu that apparently does not represent the test set as well. Also, PIFu normalizes every scan at training and prediction time to have its geometric center at the coordinate origin, whereas we use origin placed scans with slight displacements. Lastly, PIFu performs a size normalization of the body using the initial 3D body configuration estimate. The image is rescaled so that the height of the person matches the canonical size. This makes person height estimation for PIFu impossible, whereas we properly reconstruct it---at the cost of a more difficult task to solve. The benefit of this operation is not reflected in the scores because the metrics are calculated in the original image space.

When adding SemDF, we see a substantial gain in performance compared to our own baseline, but also to the so far best-performing PIFu metrics. We outperform PIFu on average with an improvement of over 50\% on the RenderPeople dataset and an average improvement of over 60\% on the BUFF dataset. When adding the Granular Render-and-Compare loss, these number improve again slightly, especially on the normal estimation. Additionally, the results gain a lot of visual fidelity and we manage to remove a lot of visual artifacts.

\begin{figure}
\centering
\includegraphics[width=\linewidth]{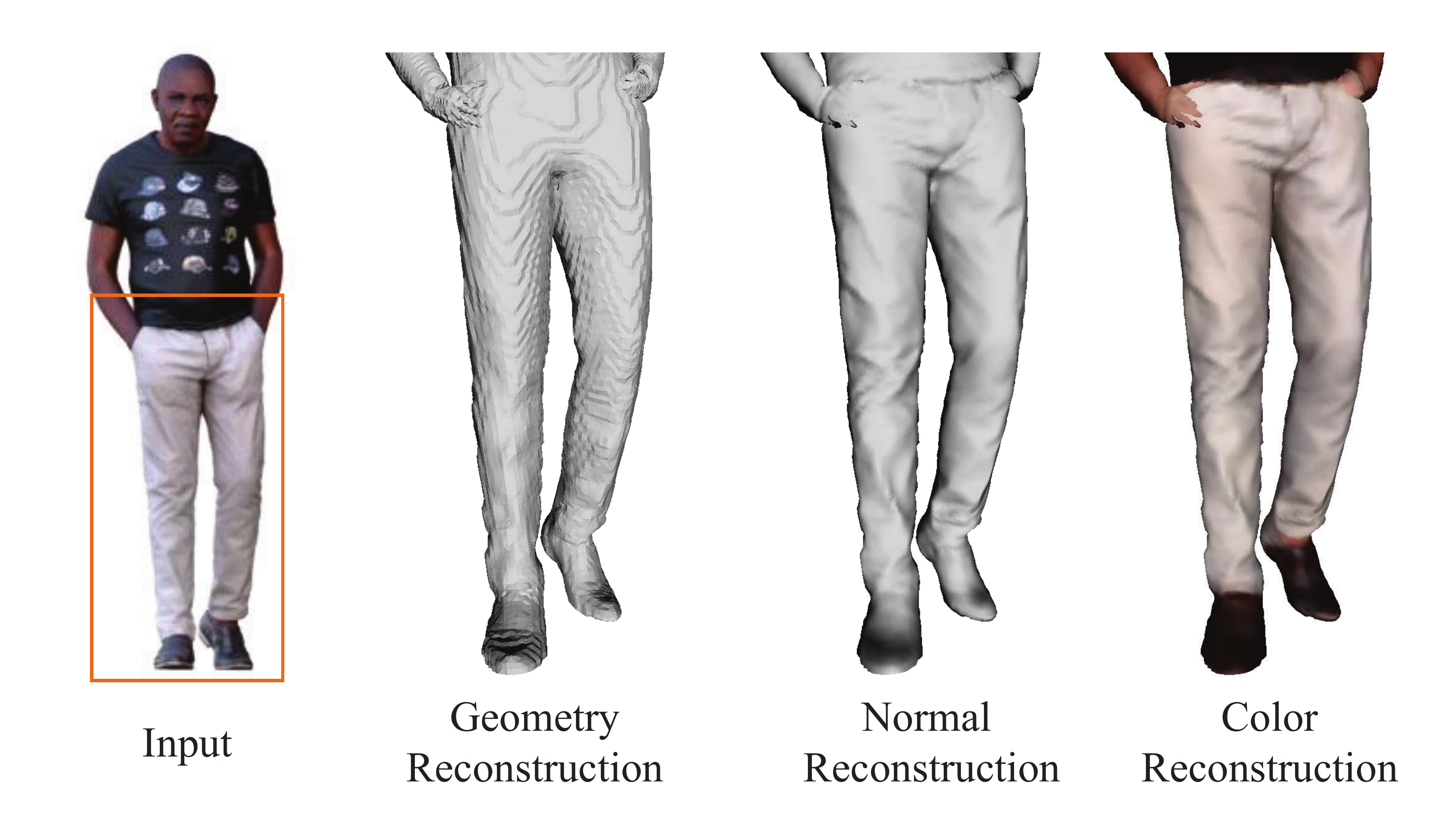}
\beforefigcaption
\caption{\textit{Reconstruction quality of clothing details.} The geometry reconstruction from our method reproduces larger wrinkles and the seam of the pants and shoes while the predicted normals reproduce fine wrinkles. The normal and color predictions rendered together produce a plausible image.}
\afterfigcaption
\label{fig:local}
\end{figure}

\begin{figure*}
\centering
\includegraphics[width=\textwidth]{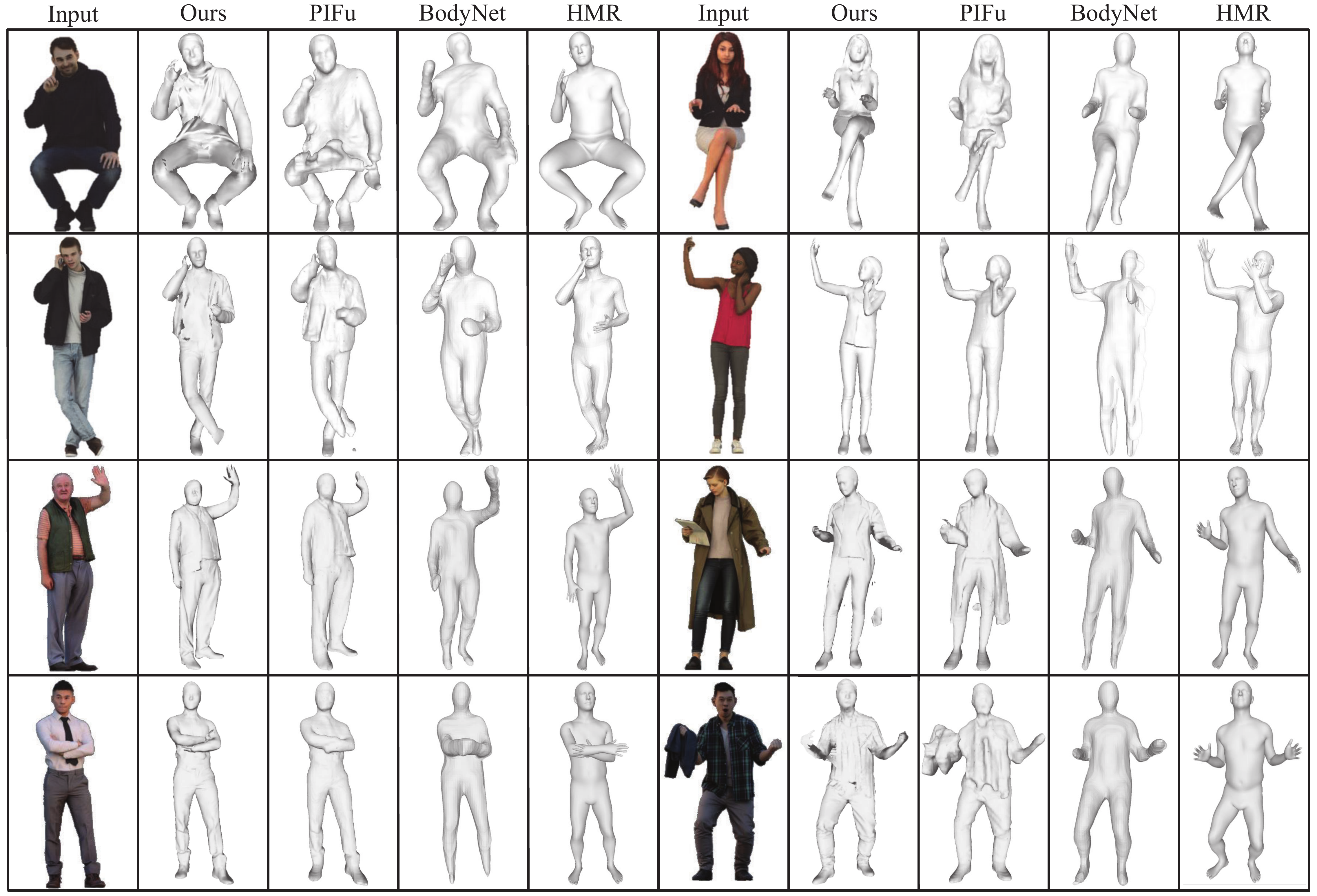}
\beforefigcaption
\caption{\textit{Qualitative comparisons against state-of-the-art methods~\cite{kanazawa2018hmr,varol18_bodynet,PIFuICCV19} on unseen images.} ARCH (Ours) handles arbitrary poses with self-contact and occlusions robustly, and reconstructs a higher level of details than existing methods. Images are from RenderPeople.
Results on DeepFashion are of similar quality but are not shown due to copyright concerns. Please contact us for more information.}
\afterfigcaption
\label{fig:result_table}
\end{figure*}

Fig.~\ref{fig:local} shows the level of detail of geometry, normal and color predictions our model can achieve. Note that, for example, the zipper is not reproduced in the predicted normal map. This is an indicator that the model does not simply reproduce differences in shading directly in the normal map, but is able to learn about geometric and shading properties of human appearance.
In Fig.~\ref{fig:BUFF}, we show qualitative results on challenging poses from the BUFF dataset.
In Fig.~\ref{fig:result_table}, we provide a comparison of results of our method with a variety of state of the art models~\cite{varol18_bodynet,kanazawa2018hmr,PIFuICCV19}.

\begin{figure}
\centering
\includegraphics[width=\linewidth]{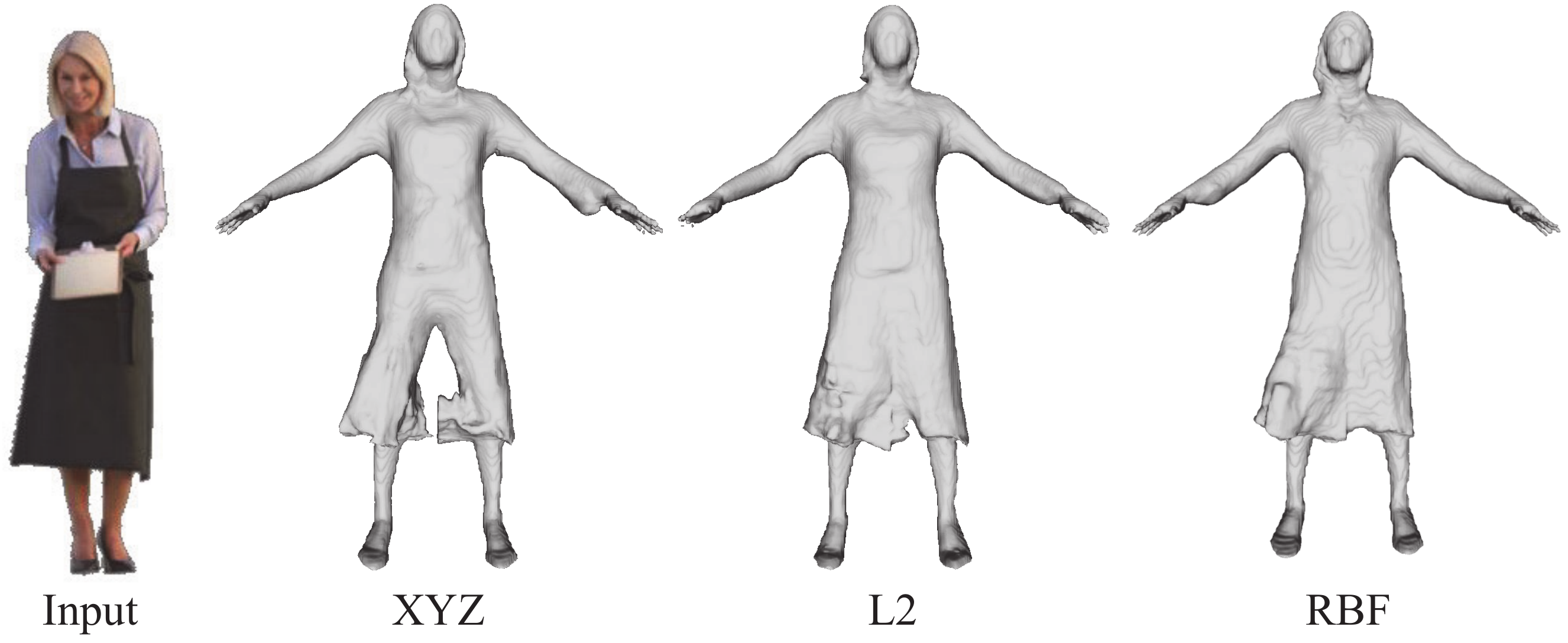}
\beforefigcaption
\caption{\textit{Reconstruction example using different types of spatial features.} \textbf{XYZ}: absolute coordinates, 
\textbf{L2}: Euclidean distances to each joint, \textbf{RBF}: Radial basis function based distance to each joint. The proposed RBF preserves notably more details.}
\afterfigcaption
\vspace{3mm}
\label{fig:ablation}
\end{figure}

\begin{table}
\centering
\setlength{\tabcolsep}{12pt}
\renewcommand\arraystretch{1.1}
\resizebox{\linewidth}{!}{
\begin{tabular}{c||c|c|c}
\hline\thickhline
Spatial Feature Types  & Normal     & P2S     & Chamfer   \\ 
\thickhline
XYZ    & 0.045     & 0.75     & 0.91   \\ 
L2     & 0.043     & 0.76     & 0.89   \\ 
RBF    & \bf{0.042} & \bf{0.74} & \bf{0.85} \\
\hline\thickhline
\end{tabular}}
\beforetab
\caption{\textit{Ablation study on the effectiveness of spatial features.} The XYZ feature uses the plain location of body landmarks. The L2 and RBF features both improve the performance.}
\aftertab
\label{tab:ablative}
\end{table}

\textbf{Ablative Studies}. We evaluate the effectiveness of different types of spatial features in Tab.~\ref{tab:ablative} and Fig.~\ref{fig:ablation}. We evaluate three different features: \textbf{XYZ} uses the absolute position of the sampled point, \textbf{L2} uses the Euclidean distance from the sampled point to each body landmark, and \textbf{RBF} denotes our proposed method in Sec.~\ref{sec:bdf}. It can be observed that RBF feature works best for this use case both qualitatively and quantitatively. RBF features strongly emphasize features that are close in distance to the currently analyzed point and puts less emphasis on points further away, facilitating optimization and preserving details.

\textbf{Animating Reconstructed Avatars}. With the predicted occupancy field we can reconstruct a mesh that is already rigged and can directly be animated. We show the animation of an avatar we reconstructed from the AXYZ dataset in Fig.~\ref{fig:animation}, driven by an off-the-shelf retargetted Mixamo animation~\cite{DenseRaCICCV19}. By working in the canonical space, the avatar is automatically rigged and can be directly animated. Given only a single view image, the avatar is reconstructed in 3D and looks plausible from all sides.

\begin{figure}[ptb]
\centering
\includegraphics[width=0.95\linewidth]{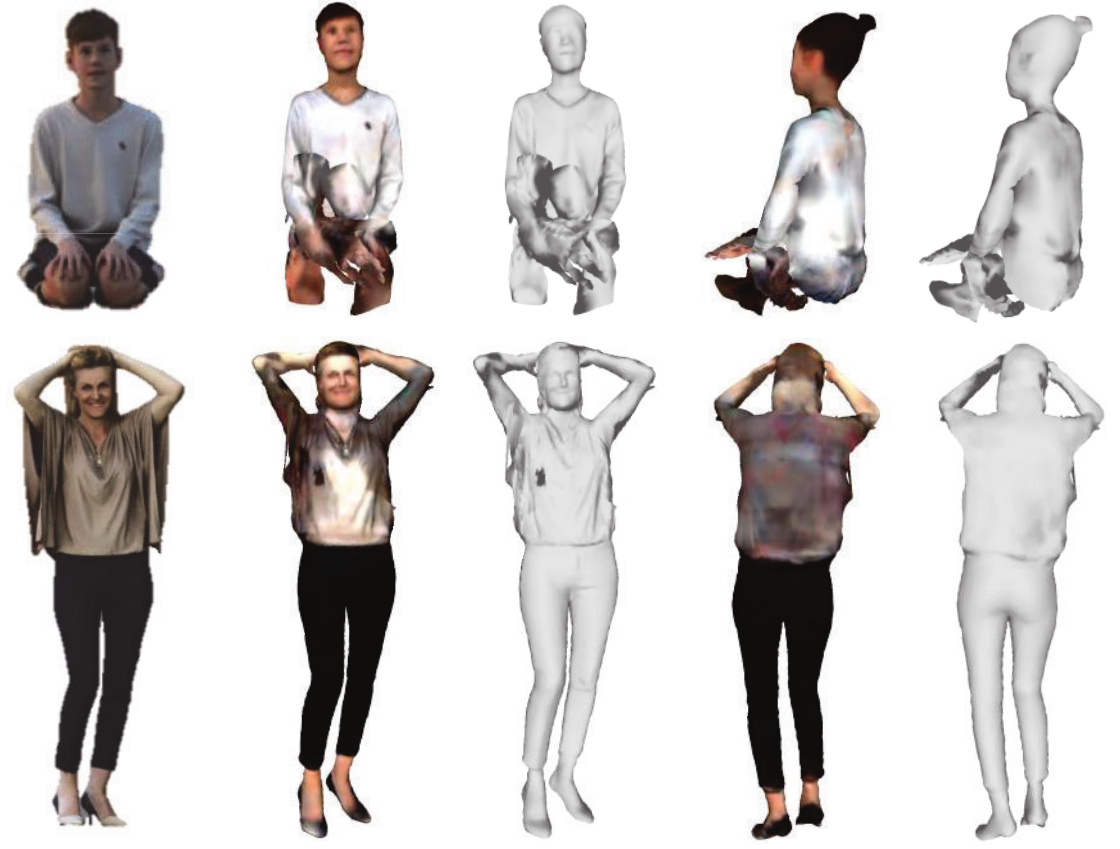}
\beforefigcaption
\vspace{3mm}
\caption{\textit{Challenging cases}. Reconstruction of rare poses, and details in occluded areas could be further improved.}
\afterfigcaption
\label{fig:fail}
\end{figure}

As shown in Fig~\ref{fig:fail}, rare poses not sufficiently covered in the training dataset (\eg, kneeing) return inaccurate body prior, and are then challenging to reconstruct.
Also, details (i.e., normals) in occluded areas could be improved with specific treatment of occlusion-aware estimation. 

\beforesection
\section{Conclusion} \label{sec:conclusion}
\aftersection

In this paper, we propose ARCH, an end-to-end framework to reconstruct clothed humans from unconstrained photos. By introducing the Semantic Space and Semantic Deformation Field, we are able to handle reconstruction from arbitrary pose. We also propose a Granular Render-and-Compare loss for our implicit function representation to further constrain visual similarity under randomized camera views.
ARCH shows higher fidelity in clothing details including pixel-aligned colors and normals with a wider range of human body configurations. The resulting models are animation-ready and can be driven by arbitrary motion sequences. 
We will explore handling heavy occlusion cases with in-the-wild images in the future.\\

\noindent {\small{\textbf{Acknowledgements.}} We would like to thank
Junbang Liang and Yinghao Huang (Interns at FRL) for their work
on dataset creation.}

{\small
\bibliographystyle{ieee_fullname}
\bibliography{3d_human}
}

\end{document}